\documentclass[12pt,a4paper]{article}

\usepackage{geometry}
\usepackage[T1]{fontenc}
\usepackage{amsmath,amssymb,amsfonts}
\usepackage{graphicx}
\usepackage{setspace}
\usepackage{mathpazo}
\usepackage{rotating}
\usepackage{float}
\usepackage{tikz}
\usepackage{xcolor}
\usepackage{subcaption}
\usepackage[colorlinks=true,allcolors=blue]{hyperref} 

\usepackage{verbatim}

\newtheorem{theorem}{Theorem}

\newtheorem{definition}{Definition}
\newtheorem{example}{Example}

\newtheorem{remark}{Remark}

\usepackage{booktabs,tabularx,multirow}
\newcolumntype{x}{>{\centering\arraybackslash}X}
\usepackage{dcolumn} 
  \newcolumntype{d}[1]{D{.}{.}{#1}}

\newcommand{\KS}{{\rm KS}}
\newcommand{\KL}{{\rm KL}}
\newcommand{\CvM}{{\rm CvM}}
\newcommand{\W}{{L_1}}

\newcommand{\E}{{\rm \bf E}}

\newcommand{\calG}{{\cal G}}

\newcommand{\calP}{{\cal P}}

\usepackage{pgfplots}
\pgfplotsset{compat=1.18}

\usepackage[textwidth=30mm]{todonotes}

\usepackage{natbib}
\usepackage{bibentry}

\newcounter{figurecounter}
\begin{document}

\title{\textbf{Adversarial Selection%
\thanks{We thank Ron Solan for useful discussions,
Tamar Itzkovitz for valuable research assistance,
and Geoffroy de Clippel, Jack Fanning, Faruk Gul, Teddy Mekonnen, Fedor Sandomirskiy, Roberto Serrano, and Leeat Yariv, as well as seminar audiences in Berlin, Brown, EUI, Princeton, and Tel Aviv for their comments.
Solan acknowledges the support of the Israel Science Foundation, Grant \#211/22.}
}

\author{Alma Cohen\thanks{Harvard Law School and Berglas School of Economics, Tel Aviv University.} \qquad Alon Klement\thanks{Buchmann Faculty of Law, Tel Aviv University.} \qquad  Zvika Neeman\thanks{EUI and Berglas School of Economics, Tel Aviv University.} \qquad Eilon Solan\thanks{School of Mathematical Sciences, Tel Aviv University.} }}

\date{March 7, 2026}

\maketitle

\begin{abstract}

In many institutional settings, $k$ items are selected with the goal of representing the underlying distribution of claims, opinions, or characteristics in a large population. We study environments with two adversarial parties whose preferences over the selected items are commonly known and opposed. We propose the Quantile Mechanism: one party partitions the population into $k$ disjoint subsets, and the other selects one item from each subset. We show that this procedure is optimally representative among all feasible mechanisms, and illustrate its use in jury selection, multi-district litigation, and committee formation. [100 words]


\end{abstract}

\bigskip
\textsc{keywords}: Nash implementation, cut-and-choose, Jury selection, MDL, redistricting committees.


\thispagestyle{empty}
\newpage
\setcounter{page}{1}

\section{Introduction}

In many institutional settings, a decision maker must select a small subset from a large population. 
The selected subset is intended to be representative of the underlying distribution of preferences, claims, opinions, or characteristics in the population. 
Yet the size of the subset is typically constrained by legal, administrative, or practical considerations, ruling out reliance on the large-sample guarantees of conventional statistical sampling theory. 
The central challenge is therefore institutional rather than purely statistical: how to design a selection procedure that produces a small sample whose empirical distribution closely tracks that of the full population. 

We study this problem in environments in which two adversarial parties possess ordinal information about the population and have opposing preferences over the composition of the selected subset. 
This structure arises in several important contexts, including: (1) jury selection from a pool of eligible jurors; (2) Multi-District Litigation (MDL), where the goal is to identify a representative sample of legal cases filed against a single defendant, out of a pool of cases filed and consolidated in one court; and (3) the appointment of members to independent redistricting commissions.\footnote{Other authors have examined this problem in the context of the selection of arbitrators, panels of judges, and members of citizen assemblies. See the references below.}

In each case, a small group must be selected from a population that can be ranked according to its relative support for one of two parties, and the parties disagree over which individuals or cases should be chosen. The problem is how to design a selection procedure that harnesses the parties' information to produce a sample whose composition best reflects the distribution of opinions or characteristics in the population.

Formally, we consider a population that consists of $n$ ranked items, but the ranking is unknown. 
Two players observe the ranking and have opposing preferences over the items. 
One player prefers higher ranked items, while the other player prefers lower ranked items. 
The goal is to select a representative sample of size $k$.




The problem can be viewed as one of design behind a veil of ignorance and can be modeled as a Nash implementation problem \citep{Maskin1999}. 
The rules governing the interaction must be chosen \emph{ex ante}, before the parties' preferences are known, while during the interaction preferences are assumed to be common knowledge. The design objective is therefore to select rules that perform well for every possible realization of preferences.

For several reasons, the problem we consider is both tractable and interesting from an implementation perspective. First, the applications we consider all involve only two parties, who are engaged in playing a zero-sum game. Thus, it is
reasonable to assume they have completely opposed or nearly completely opposed rankings over the population.
Second, the applications we consider are all cases where  the assumption of complete information, which underlies our theoretical analysis, is reasonable. Indeed, in all three applications, the involved parties share the same information with respect to the individuals or items to be selected. 
Third, in all three applications, the goal is to select a small sample, which is representative of the entire population.

We evaluate the ``representativeness'' of the chosen sample by the distance between the cumulative distribution generated by a sampling procedure and that of the full population. 
We propose a selection rule, which we call the \emph{Quantile mechanism}, that uses only the players' ordinal ranking information to construct a sample that is maximally representative for any possible distribution of items, according to three different distance measures: Kolmogorov-Smirnov (KS), $L_1$, and the Cram\'er-von Mises statistic (CvM).

The Quantile mechanism is inspired by the cut-and-choose procedure from the cake-cutting literature \citep{BramsTaylor1996}. 
One player partitions the population of $n$ items into $k$ disjoint sets, and the other player selects one item from each set.\footnote{Throughout, “partition” refers to a collection of disjoint sets; these sets need not cover the entire population.} 
The Quantile mechanism builds on the observation that the player who partitions the population, say it is the player who prefers high-ranking items, anticipates that the other player, who prefers low-ranking items, will choose the lowest-ranked item from every set. 
So the first player has an incentive to place all the highest-ranking items into the smallest set, the next high-ranking items into the next smallest set, and so on. 
Similarly, if the player who prefers low-ranking items is chosen to partition the population, then she has the opposite incentive to place all the lowest-ranking items into the smallest set, the next low-ranking items into the next smallest set, and so on. 
Importantly, the Quantile mechanism is symmetric: it yields the same set of choices, or chosen sample, regardless of whether the partitioning is performed by the first or second player.

Our main result is that the \textit{Quantile} mechanism produces the most representative sample under all three distance measures mentioned above. 
In particular, the Quantile mechanism is better than all the mechanisms that are currently used in the settings mentioned above, including random selection and the strike-out mechanism used in jury selection and in MDLs.\footnote{
The Quantile mechanism is also superior to the direct mechanism in which both players report the ranking. If the reports coincide, the mechanism selects the best possible sample given the reported ranking (which as our analysis shows selects the same quantiles as those chosen by the Quantile mechanism); if they differ, it draws a random sample of size $k$. This is because truthful reporting is not a Nash equilibrium of the direct mechanism if the players preferences are diametrically opposed to each other: if one player strictly prefers to report truthfully, then the other player prefers not to, and vice-versa. So the direct mechanism can only implement a selection rule that both players find indifferent to random selection. }
Moreover, we show that the sample obtained by the Quantile mechanism is \textit{strictly} better than \textit{any} other sample of the same size.

The rest of the paper proceeds as follows. In the next section we elaborate on the three real-world settings mentioned above: jury selection, multi-district litigation case selection, and the selection of independent redistricting commissions. Section \ref{sec:literature} surveys the related theoretical literature. Section \ref{sec:model} sets up the model. Section \ref{sec:distance} describes three alternative ``representativeness''  statistics: Kolmogorov-Smirnov (KS),  $L_1$, and the Cram\'er-von Mises statistic (CvM), and presents the main result. Section \ref{sec:discussion} contains three  extensions of our main result. The first result shows that any selection of quantiles can be selected by some cut-and-choose mechanism. The second result extends the main result to the case in which the two players do not have opposing preferences, and the third result addresses the case in which there is only one player. Finally, Section \ref{sec:conclusion} provides concluding remarks. All proofs are relegated to the Appendix.

\section{Application to Jury Selection, MDL, and Redistricting Committees}

In this section, we examine three different settings in which the challenge of choosing a small, representative subset from a large population arises: jury selection, Multi-District Litigation, and the appointment of members to independent redistricting commissions. We briefly outline how selection is currently carried out in each of these settings. As we demonstrate in the following sections, in all of these settings, substituting existing procedures with the Quantile mechanism would enhance the ``representativeness'' of the individuals or items selected.

\bigskip

\noindent\textbf{\textsc{Jury Selection.}}
Pursuant to 28 U.S.C. \S 1863, each United States district court is required to formulate and implement a written plan governing the random selection of jurors. This plan must be designed to ensure the selection of a fair cross section of the individuals residing within the community of the district or division in which the court is located. The fair cross section requirement derives from the Sixth Amendment to the United States Constitution, as construed by the Supreme Court in \textit{Taylor v. Louisiana}, 419 U.S. 522 (1975), and\textit{ Duren v. Missouri}, 439 U.S. 357 (1979).  

The jury selection process begins with the compilation of a master list. 
 From this master list, a random subset of eligible prospective jurors is drawn and summoned to court for further evaluation.  The group of qualified individuals from which the final jury is selected is termed the \textit{venire}
 \citep{kovera2013jury}. 
In the selection process the plaintiff (or, in criminal cases, the prosecution) and the defendant may request that specific jurors be excluded ``for cause.'' In addition, each party is allotted a certain number of \textit{peremptory} challenges, which allow them to strike jurors without stating a cause, with the dismissed jurors replaced by other prospective jurors who were summoned for that same day 
\citep{wagner1981art}.\footnote{The specific procedure used to let the parties exercise their challenges varies greatly across jurisdictions and is sometimes left to the discretion of the judge. Two classes of procedures
are most frequently used. In \emph{Struck} procedures, the parties can observe and extensively question all the jurors who could potentially serve on their trial before
exercising their challenges (this questioning process is known as voir dire). In contrast, in
\emph{Strike and Replace} procedures, smaller groups of jurors are sequentially presented to the parties. The parties observe and question the group they are presented
with (sometimes a single juror) but must exercise their challenges on that group without knowing the identity of the next potential jurors \citep{moro2024exclusion}.}


Empirical evidence consistently indicates that minority groups are underrepresented in jury pools and venires. For example, \citet{rose2018jury} document pervasive minority underrepresentation in federal jury pools, with typical “missing” minority counts in venires that are modest in absolute magnitude but substantial in probabilistic terms for determining whether any minority jurors are present. In a related vein, \citet{anwar2022unequal} show that geographic disparities in representation among seated juries closely track disparities already present in the pools of potential jurors.  

Misrepresentation of certain groups is also documented with respect to seated juries. \cite{AnwaretalQJE2012} find that the reduced-form effect of pool composition on case outcomes is substantially larger than what naive correlations based solely on the race of seated jurors would suggest, consistent with the interpretation that attorney selection behavior and other strategic responses confound comparisons based on seated-jury characteristics.
\citet{anwar2014role} further show that prosecutors disproportionately strike younger potential jurors, whereas defense attorneys disproportionately strike older potential jurors, resulting in reduced seating probabilities at both tails of the age distribution relative to its center. \citet{flanagan2018race} reports that the demographic composition of the randomly selected pool causally affects conviction probabilities and that attorneys adapt their peremptory strategies in response. By contrast, \citet{diamond2009achieving} find that opposing peremptories may partly offset one another on average, while structural features such as jury size exert an independent and more stable influence on minority representation.

\bigskip
\noindent\textbf{\textsc{Multi-District Litigation.}}
A Multi-District Litigation (MDL) is a federal procedural mechanism created under 28 U.S. Code \S 1407. It allows civil lawsuits that are pending in different federal districts, but share common factual questions, to be centralized in a single district court (the \textit{transferee} court) for coordinated pretrial proceedings.  MDL cases consist of a significant portion of the federal docket, and at times have accounted for more than 40 percent of all federal civil cases \citep{gluck2021mdl}. 
Although the transferee court - where all cases are centralized - is in principle responsible only for overseeing pretrial proceedings, in practice more than 98 percent of these cases end there, typically through settlement or dismissal \citep{fallon2020bellwether}.

Because both the parties and the court need reliable information about the likely value of MDL claims, yet cannot practically conduct a full trial for each one, judges often turn to a \textit{bellwether} trial process to generate these estimates. The label “bellwether” derives from the practice of placing a bell on the lead sheep (a wether) to guide the flock, underscoring that these trials are meant to guide resolution of the remaining cases. The outcomes of bellwether trials do not have preclusive or binding effect on litigants who are not parties to those particular trials and instead serve only as informational inputs. The value of this approach, however, hinges on whether the selected cases are truly representative of the broader pool of claims \citep{fallon2007bellwether,lahav2007bellwether,lahav2018primer,whitney2019bellwether}.

Transferee judges employ bellwether trials to promote efficient resolution of complex MDL dockets. These trials serve to collect information about liability, causation, and defenses in an actual trial setting; to illuminate how juries respond to the evidence and expert testimony; to establish benchmarks or reference points for valuing the remaining cases in settlement negotiations --
often prompting global settlements; and to assist courts in resolving recurring legal and evidentiary questions.

The process of selecting bellwether cases is therefore crucial, since it  hinges on the representativeness of the cases chosen. Courts use a variety of  techniques for selecting the bellwether cases, generally grouped into party selection, random selection, and judicial selection. Under a party selection approach, plaintiffs and defendants each choose a designated number of cases. Alternatively, cases can be drawn at random, either from the general pool of cases, or through stratified sampling. Finally, the court may select a representative sample, often after receiving recommendations from the parties and using some random sampling techniques \citep{whitney2019bellwether, fallon2020bellwether}.

Although the legal literature articulates substantial concerns regarding the representativeness of bellwether trials selected through the aforementioned techniques, systematic empirical assessment of these selection processes remains limited. \citet{brown2014bellwether} analyze the cases chosen in the Vioxx multidistrict litigation (MDL) proceedings and demonstrate that cases selected by plaintiffs’ representatives deviate significantly from what would be expected under random selection, whereas no comparable divergence is observed for cases selected by defendants. These results, however, are derived from a single MDL proceeding. \citet{villalon2022different} compares case-selection practices across two MDLs and concludes that, as the size of the consolidated pool increases, judges exhibit a stronger preference for random selection over party-driven selection.

\bigskip
\noindent\textbf{\textsc{Selection of Independent Redistricting Commissions.}}
Redistricting is the process of redrawing the lines of legislative districts, grouping voters into geographic territories from which they elect their representatives. This process affects districts at all levels of government, including local councils, state legislatures, and the U.S. House of Representatives. The way these lines are drawn determines which voters are grouped together, directly influencing which communities are represented and which candidates can win elections.

The redistricting process typically begins following the federal Census, which is conducted every ten years at the start of a new decade. The legal mandate for this cycle is rooted in the ``one person, one vote'' constitutional principle established by the U.S. Supreme Court in the 1960s. In cases such as \textit{Reynolds v. Sims} (377 U.S. 533, 1964) and \textit{Wesberry v. Sanders} (376 U.S. 1, 1964), the Court ruled that legislative districts must contain roughly equal populations. 

Because populations shift over time, states are practically required to redraw district lines after every Census to ensure this equality is maintained \citep{karlan1992rights}. While lines must be redrawn at least once per decade, some states allow for redistricting to occur more frequently, whereas others prohibit mid-decade change \citep{cox2004partisan}.

States use different procedures to determine who draws electoral district boundaries. When legislators draw their own district lines, a conflict of interest arises: politicians may effectively choose their voters rather than voters choosing their representatives. Incumbents may therefore engage in gerrymandering, or the manipulation of district boundaries to influence electoral outcomes \citep{imamura2022rise}.

To mitigate this concern, some states, such as Colorado, 
Michigan, and California, delegate redistricting to independent citizen commissions.\footnote{Colorado Independent Redistricting Commissions, Commissioner Selection Process, \href{https://redistricting.colorado.gov/content/commissioner-selection-process}{https://redistricting.colorado.gov/content/commissioner-selection-process}; Jason  Torchinsky and Dennis W. Polio, \textit{How Independent is Too Independent?: Redistricting Commissions and the Growth of the Unaccountable Administrative State, }20 Geo. JL and Pub. Pol’y 533,\textit{ }543-50 (2022).} Members of these commissions are selected through procedures that combine random selection with multi-stage vetting by nonpartisan bodies (such as state auditors or judicial panels). These procedures typically impose constraints on the commission's partisan composition, for example, equal representation of Democratic and Republican members together with a group of independent or unaffiliated commissioners (\citet{torchinsky2022independent}). The goal is to create a commission that reflects the state's political and demographic diversity and is therefore more likely to produce fair district maps.

\section{Related Literature}\label{sec:literature}

A relatively limited body of theoretical literature examines problems closely related to those addressed in the present study. \citet{deClippel_AER2014} analyze the selection of an arbitrator as a Nash implementation problem. 
They propose a procedure they call \emph{shortlisting}, in which one party selects a subset of $\frac{n+1}{2}$ items out of $ n$  and the other party chooses one item from this subset. 
They show that this mechanism implements a Pareto-efficient outcome that Pareto-dominates the median alternative for both parties.\footnote{See \citet{BarberaCoelhoRAND2022} for a comparison of variants of this procedure that allow the parties to determine the role of proposer and the size of the shortlist. See \citet{BarberaCoelhoBSEWP2024} for generalizations of this procedure to the selection of $k \geq 1$ alternatives in the context of the selection of arbitrator and judge panels.} 

When $k=1$, their shortlisting procedure coincides with our Quantile mechanism. 
However, our contribution differs along several dimensions. 
First, we study the selection of $k \geq 1$ items rather than a single arbitrator. 
Second, their objective is Pareto efficiency, whereas our focus is on representativeness, measured by the distance between the sample and population distributions. 
(Importantly, the Quantile mechanism is also Pareto efficient in our setting.) 
Third, they allow for general preference profiles, while we focus on environments with strictly opposed preferences.\footnote{See Section \ref{sec:antagonistic} for an extension of our framework to more general preference profiles.} 

More recently, \citet{BogomolnaiaetalTE2024} have considered a more general version of this problem with possibly more than two agents who need to choose one alternative from a given set. They show that any mechanism guaranteeing a maximal welfare level to the agents must either combine variants of random dictatorship and voting-by-veto mechanisms or be purely random, depending on the parameters of the model. 
Their objective is welfare maximization, whereas our criterion is distributional representativeness.


There is also less directly related literature in theoretical computer science which studies the design of representative citizens' assemblies under demographic constraints. 
See, for example, \citet{Flanigan2021} and references therein. 
The papers in this literature aim to maximize individuals' selection probabilities subject to the constraint that different (observable) demographic groups are fairly represented in the assembly. 
In contrast, we consider a setting in which the planner does not observe the underlying ranking of items and must rely solely on the ordinal information held by the parties.


\section{Model}\label{sec:model}

A \emph{population} consists of $n$ items
$x = (x_1,\dots, x_{n})$.
The items are ranked according to a complete and transitive ranking $\succsim$.
A social planner,
who 
holds no information about the ranking,
would like to select a sample of size $k$, which is representative  of the population, in a way we define below. 
The social planner can solicit help from two players,
I and II, 
who are aware of the ranking, but have their own different preferences over the selected sample:
Player I prefers items whose ranking is higher, while player II prefers items whose ranking is lower. 

A selection procedure, or a mechanism, is a game-form that selects 
(possibly randomly) a set of $k$ indices. 
The mechanism specifies a message set for each one of the players, together with a mapping from the players' chosen messages into 
sets of $k$ indices (if the mechanism is deterministic)
or into distributions over sets of $k$ indices (if the mechanism is random). 
The mechanism induces a normal-form game in which the players, who are commonly known to know the ranking of the items, choose their messages simultaneously, trying to induce the selection of highly and lowly placed items according to the ranking, respectively. We focus on a Nash-equilibrium of this game. 



Fixing the population size $n$ and the desired sample size $k \leq n$, denote by 
$\calP(n,k)$
the collection of all subsets of indices $\{1,\dots,n\}$ that have size $k$.

A mechanism is formally defined as follows:


\begin{definition}[Mechanism]
A \emph{mechanism} is a tuple 
$M = \langle A_I,A_{II},f\rangle$ where
\begin{itemize}
\item $A_I$ is a finite set of messages for Player~I,
\item $A_{II}$ is a finite set of messages for Player~II,
\item $f$ is a function that,
for every pair of actions $a_I \in A_I$ and $a_{II} \in A_{II}$,
returns a probability distribution over $\calP(n,k)$.
\end{itemize}
\end{definition}

\begin{remark}
We describe mechanisms as simultaneous-move game forms. This representation is without loss of generality. Sequential procedures, including those involving moves by nature, can also be represented within this framework using the standard transformation from extensive-form games to normal-form games.
\end{remark}



The \emph{outcome} of a mechanism is a probability distribution over the set $\calP(n,k)$. In order to analyze how the players play in the game that is induced by a mechanism $M$, we need to describe the players' preferences over the space of probability distribution over $\calP(n,k)$.

\begin{definition}[Utility function]
Let $x = (x_1,\dots,x_{n})$ be a population,
let $\succsim$ be a ranking on $x$,
and let $y$ and $y'$ be two samples of $k$ items from $x$. A \emph{utility function} of Player~I is a function $u_I$ that assigns a real number to every sample of $k$ items,
such that $u_I(y) \geq u_I(y')$ whenever the sample \emph{$y$ is shifted to the right relative to $y'$}, or
\[ \#\{j \in \{1,\dots,k\} \colon y_j \precsim x_i\} \leq
\#\{j \in \{1,\dots,k\} \colon y'_j \precsim x_i\}, \ \ \ \forall i \in \{1,\dots,n\}. \]
Similarly, a \emph{utility function} of Player~II is a function $u_{II}$ that assigns a real number to every sample of $k$ items, such that $u_{II}(y) \geq u_{II}(y')$ whenever $y'$ is shifted to the right relative to $y$.
\end{definition}

\begin{remark}

Notice that the statement that the sample ``$y$ is shifted to the right relative to $y'$'' is equivalent to the requirement that the cumulative distribution function (CDF) of the sample $y$, defined as
\begin{equation}\label{eq:sampleCDF}
    F_y(x_i) \equiv \frac{1}{k}\cdot \#\{j \in \{1,\dots,k\} : y_j \precsim x_i\}, \ \ \ \forall i \in \{1,\dots,n\},
\end{equation}
first-order stochastically dominates the CDF of $y'$, defined analogously. 
Equivalently, the sample ``$y$ is shifted to the right relative to $y'$'' holds if and only if
\[
F_y(x_i) \leq F_{y'}(x_i) \quad \text{for every } i \in \{1,\ldots,n\}.
\]
 \end{remark}

The assumption that the players' utility functions are monotone with respect to first-order stochastic dominance reflects the assumption that Player~I prefers items that are ranked higher, while Player~II prefers items that are ranked lower.

We do not impose any additional assumptions on the comparison of two samples that are not ordered by stochastic dominance.

\begin{remark}
Although it is natural in our applications to assume that the two players' preferences are exactly opposed, the analysis below does not rely on this assumption. In particular, our results allow for preference profiles in which $u_I \neq -u_{II}$.
\end{remark}


Since the outcome of a mechanism is a probability distribution over samples, we must specify the players' utility functions on this space. We assume that the players' preferences satisfy the von Neumann-Morgenstern axioms, so that they are expected-utility maximizers. 
That is, a player's utility from a probability distribution over samples equals the expected utility of the realized sample.

The specification of a mechanism induces a complete information game between the players, as follows. 

\begin{definition}[Mechanism game]
A \emph{mechanism game} is a triplet $\langle M,u_I,u_{II}\rangle$, where $M = \langle A_I,A_{II},f\rangle$ is a mechanism, and 
$u_I$ and $u_{II}$ are the two players' payoff functions.
\end{definition}

A mechanism game is played as follows.
\begin{itemize}
\item Fix a population $x = (x_i)_{i=1}^{n}$ and a ranking $\succsim$ over $x$.
\item 
After observing $x$ and $\succsim$,
Player~I (resp., Player~II) selects $a \in A_I$ 
(resp., $a_{II}\in A_{II}$).
This selection is made simultaneously.
\item A set of $k$ indices $S \in \calP(n,k)$ is selected randomly according to the probability distribution $f(a_I,a_{II})$.
\item The outcome of the game is the vector $y := (x_s)_{s \in S}$.
\end{itemize}

A \emph{pure strategy} for Player~I is a function 
that assigns a message in $A_I$ to any possible input $x, \succsim$,
and a \emph{mixed strategy} for Player~I is a function
that assigns a probability distribution on $A_I$ to any possible input $x, \succsim$.
Pure and mixed strategies for Player~II are defined analogously.


We assume that the players play a Nash equilibrium of the induced game, which is defined as follows:

\begin{definition}[Equilibrium]
A pair of strategies $(\alpha_I^*,\alpha_{II}^*)$ is a \emph{Nash equilibrium}
if for every two strategies $\alpha_I$ of Player~I and $\alpha_{II}$ of Player~II we have
\begin{align*}
\E_{\alpha_I^*,\alpha_{II}^*}[u_I(y)] &\geq \E_{\alpha_I,\alpha_{II}^*}[u_I(y)],
\ \ \ \forall \alpha_I,\\
\E_{\alpha_I^*,\alpha_{II}^*}[u_{II}(y)] &\geq \E_{\alpha_I^*,\alpha_{II}}[u_{II}(y)],
\ \ \ \forall \alpha_{II}.
\end{align*}
\end{definition}

By Nash's Theorem,
every mechanism game admits a Nash equilibrium in mixed strategies.

Note that there are two sources of randomization in the play. First, the players select their messages at random,
and second, the sample is selected randomly according to $f$. A mechanism game may well have a unique Nash equilibrium in mixed strategies. 

We next provide some examples of selection mechanisms that illustrate the model and how mechanisms can be used to select representative samples.
The first example shows how it is possible to implement the selection of a random sample from the population.

\begin{example}[Random Sample]
\label{example:uniform}
Fix a population $x = (x_i)_{i=1}^{n}$. The mechanism $M = \langle A_I,A_{II},f\rangle$ that implements the selection of a random sample of size $k$ from the population $x$ is defined as follows: $A_I$ and $A_{II}$ are arbitrary sets. In particular, they can be singletons.
The function $f(a_I,a_{II})$ is the uniform distribution over $\calP(n,k)$,
for every $a_I \in A_I$ and $a_{II} \in A_{II}$.
Since the players' messages do not affect the outcome of the mechanism, any pair of strategies is an equilibrium of the corresponding mechanism game.
\end{example}

\begin{example}[Strike and Replace Mechanism]
\label{example:struck_jury} 
Fix a population $x = (x_i)_{i=1}^{n}$, a ranking $\succsim$ over $x$, and a sample size $k\leq n$. The procedure selects a random sample of size $k$, lets player $I$ veto up to $c$ items, and then lets plater $II$ veto up to $c$ items. Each vetoed item is replaced with a freshly drawn item from the population, without replacement. The procedure outputs the sample that survived this process of elimination. This procedure can be implemented by a mechanism in which the two players' message sets $A_I$ and $A_{II}$ include all the possible sequential veto strategies.

In equilibrium, each player will veto the $c$ items she dislikes the most from the sample (lowest ranking item for Player I; and highest ranking item for Player II) provided these items are worse than a randomly sampled item from the population.
\end{example}

\begin{example}[Median-Sample Mechanism]
\label{example:median_sample} An interesting improvement to both the Random Sample Mechanism and the Struck Jury Mechanism was recently proposed by \cite{flanagan2025not}. Flanagan suggests drawing $2c+1$ random samples, and letting each player veto $c$ of these samples. This procedure can also be implemented by a mechanism. In equilibrium, each player will veto the $c$ samples she dislikes the most, which will produce the median sample as the outcome of the mechanism. \cite{flanagan2025not} claims that this procedure reduces the variance of the selected sample relative to both random selection and the Struck Jury Mechanism.
\end{example}

The next example shows that there exists a mechanism that always outputs the median item from any population $x=(x_i)_{i=1}^{n}$ in equilibrium.

\begin{example}[The Median Mechanism]
\label{example:median}
Fix a population $x = (x_i)_{i=1}^{n}$ and a ranking $\succsim$ over $x$. Suppose that $k=1$. The mechanism works as follows:
ask Player~I to select a subset $a_I$ of $\lceil \frac{n}{2} \rceil$ items from the population $x$;
ask Player~II to select one of the items in $a_I$; select the item that was selected by Player II from Player I's selected subset. As mentioned above, the Median Mechanism is equivalent to the shortlisting procedure proposed by \cite{deClippel_AER2014}. In their procedure, one of the players shortlists $\lceil \frac{n}{2} \rceil$ items, from which the other player chooses one item.

Since Player~II wants to minimize the ranking of the selected item, she will choose the minimal (according to $\succsim$) item in $a_I$.
Anticipating that, Player~I,
who wants to maximize the ranking of the selected item,
will select the subset of items that contains the top $\lceil \frac{n}{2} \rceil$ items from $x$ (according to $\succsim$).

Formally, the mechanism $M = \langle A_I,A_{II},f\rangle$ is described as follows.
\begin{itemize}
\item $A_I$ is the set of all subsets of $\{1,\dots,n\}$
that contain $\lceil \frac{n}{2} \rceil$ items.
\item 
$A_{II}$ contains all the functions $a_{II}:A_I\rightarrow\{1,\dots,n\}$
that assign to each subset $a_I \subset \{1,\dots,n\}$ of $\lceil \frac{n}{2} \rceil$ items one of the items in $a_I$.
\item $f(a_I,a_{II})$ assigns probability one to the implementation of Player~II's choice on Player~I's choice, $a_{II}(a_I)$.
\end{itemize}
If the ranking $\succsim$ over $x$ is strict, then the mechanism game admits a unique equilibrium in which
Player~I selects the set that contains the highest $\lceil \frac{n}{2} \rceil$ items, and Player~II selects the function that assigns to each subset $a_I \subseteq \{1,\dots,n\}$ of $\lceil \frac{n}{2} \rceil$ 
the item in the set that is smallest according to $\succsim$.
Otherwise, 
there is a multiplicity of equilibria,
but the equilibrium outcome is always taken from the set of items that are equivalent (according to $\succsim$) 
to the item ranked $\lceil \frac{n}{2} \rceil$.
\end{example}

The next example describes a general class of 
mechanisms that we call \emph{cut-and-choose} mechanisms. In these mechanisms, one of the players is asked to select $k$ subsets (not necessarily disjoint) from the set of indices $\{1,\ldots,n\}$ with $n_1,\ldots,n_k$ items in each set, respectively, and the other player is asked to select one index from each subset. The mechanism outputs the $k$ indices selected by the second player. By varying the numbers of indices in each set, $n_1,\ldots,n_k$, and by varying the restrictions on the extent to which these subsets can overlap, it is possible to obtain the general class of cut-and-choose mechanisms. The Median Mechanism described in Example \ref{example:median} is an example of a cut-and-choose mechanism in which $k=1$ and $n_1=\lceil \frac{n}{2} \rceil$.

\begin{example}[Cut-and-Choose Mechanism]
\label{example:I Cut You Choose}
Fix a population $x = (x_i)_{i=1}^{n}$, a sample size $k$, and a ranking $\succsim$ over $x$. 
A cut-and-choose mechanism with $k$ disjoint subsets with sizes $n_1,\ldots,n_k$, $\sum_i^k n_i\leq n$ works as follows. Suppose, without loss of generality, that $n_1\leq \cdots\leq n_k$. One of the players is asked to partition the set of indices $\{1,\ldots,n\}$ into $k$ disjoint subsets, with $n_1,\ldots,n_k$ indices each, respectively. The other player is asked to choose one index from each one of the $k$ subsets, and the mechanism outputs the chosen $k$ indices.

Suppose that Player I is asked to select the subsets (cut), and Player II is asked to choose an index from each subset (choose). In this case, the equilibrium of the cut-and-choose mechanism (which is unique up to equivalent outcomes)
is that Player I places the indices of the $n_1$ highest ranked items in the first set, the indices of the next $n_2$ highest largest items in the second set, and so on, up to the $k$-th set, and Player II chooses the index of the lowest ranked item in every subset.

To see this, note that for player II, choosing the index of the lowest ranked item from each subset of indices chosen by Player I is a dominant strategy. 
As for Player I, observe that if there are two sets of indices $I_1$ and $I_2$ such that the lowest ranked item $x^1$ from the items indexed by $I_1$ is ranked below the lowest ranked item $x^2$ indexed by $I_2$, 
that is, $x^1 \prec x^2$, 
and if there is another item $x_1'$ in $I_1$ that is ranked above $x^2$, then Player I would benefit from switching between the two items $x_1'$ and $x_2$, because the switch does not affect Player II's choice from $I_1$, and weakly increases the rank of the item chosen from the set $I_2$. This argument implies that the $k$ subsets of indices that are selected by Player I are ordered in the sense that the lowest ranked item in the $i$-th subset of indices is ranked above the highest ranked item in the subset $i+1$-th subset. 

Moreover, by putting the $n_1$ highest ranked items in the first set, the indices of the next $n_2$ highest largest items in the second set, and so on, Player I shifts the chosen sample as much to the right as possible, which is to his benefit.

It follows that in the equilibrium of this cut-and-choose mechanism the outcome contains the $n-n_1+1$-th highest ranked, $n-n_1-n_2+1$-th highest ranked item, and so on. If instead Player II was chosen to cut and Player I was chosen to choose, in equilibrium the outcome would have contained  the $n_1$-th lowest ranked, $n_1+n_2$-th lowest ranked item, and so on, up to the $\sum_i^k n_i$-th lowest ranked items from the population.
\end{example}


\section{Maximizing Representativeness}\label{sec:distance}

As explained above, our objective is to identify the mechanism that produces the sample most representative of the population. In this section, we introduce three notions of representativeness, each defined as a distance between the cumulative distribution function (CDF) of the sample and that of the population: the Kolmogorov-Smirnov statistic, the $\W$-statistic, and the Cramér-von-Mises statistic. These statistics are widely used in statistical applications.\footnote{An alternative measure of distance between distributions that is often used in the economics literature is \emph{Kullback-Leibler (KL) divergence}. KL divergence compares the population distribution (rather than CDF) to the sample distribution.
Denoting
$P_x(x_i) = \#\{j \colon x_j=x_i\}$
and
$P_y(x_i) = \#\{j \colon y_j=x_i\}$,
for each $i \in \{1,\dots,n\}$, KL divergence is given by
\[ \KL(x,y) := \frac{1}{n}\sum_{i=1}^n \log\left( \frac{P_x(x_i)}{P_y(x_i)}\right). \]
There are two reasons \emph{not} to use this measure in our setup.
First, when the number of equivalence classes of items in the population is larger than $k$,
necessarily there will be $i \in \{1,\dots,n\}$ such that $F_y(x_i) = 0$,
in which case KL divergence is not well-defined.
Second, 
even if the number of equivalence classes of items in the population is at most $k$,
the sample that minimizes KL divergence need not represent the population.
For example,
consider a population that consists of two types of items, red and blue.
Suppose that the population consists of one red item and $99$ blue items, so that $n=100$, 
and suppose that $k=2$.
The sample that minimizes the KL divergence is the one that contains one item from each type. 
However, the sample that contains two blue items better represents the population.

Another common measure of distance between distributions is  \emph{$p$-Wasserstein distance}.
This measure is applicable when there is a natural distance between items in the population, which is not the case in the applications we consider.
Under
the following natural distance between items:
the distance between items $x_i$ and $x_j$ is $0$ if $x_i \sim x_j$,
and $1$ plus the number of items that lie strictly between $x_i$ and $x_j$ otherwise,
if $p=1$, then
this distance measure is equivalent to $\W$-statistic, and hence 
the Quantile mechanism is also optimal according to this measure.}
We show that the Quantile mechanism, which is defined below, produces the most representative sample according to all three measures.

For any population $x=(x_i)_{i=1}^{n}$ and ranking $\succsim$ on $x$, denote by $F_x$ the cumulative distribution function (CDF) of $x$:
\begin{equation}\label{eq:populationCDF}
  F_x(x_i) \equiv \frac{1}{n}\cdot \#\bigl\{ j \in \{1,\dots,n\} \colon x_j \precsim x_i\bigr\}, \ \ \ \forall i \in \{1,\dots,n\}.   
\end{equation}
Similarly, 
for any sample $y = (y_i)_{i=1}^{k}$ of $k$ items from $x$,
recall the definition of the CDF $F_y$ of the sample $y$ from Equation (\ref{eq:sampleCDF}).




\bigskip
\noindent\textbf{\textsc{Kolmogorov-Smirnov Statistic.}}
The Kolmogorov-Smirnov (KS) statistic \citep{an1933sulla}  compares a population to a sample taken from that population by the \emph{maximal} difference between the sample's CDF and the population's CDF.


\begin{definition}[KS statistic]
Let $x$ be the population, 
let $\succsim$ be a ranking on $x$, 
and let $y$ be a sample of size $k$ from $x$.
The \emph{KS statistic} of $x$ and $y$ is
\[ \KS(x,y) \equiv  \max_{1 \leq i \leq n} |F_x(x_i) - F_{y}(x_i)|. \]
\end{definition}

Accordingly, we say that a sample $y$ is KS-optimal if it minimizes the KS statistic, and that a mechanism is KS-optimal if it produces the KS-optimal sample for every population.

\begin{definition}[KS-optimality]
A sample $y$ is \emph{KS-optimal} 
for the population $x$ if
\[ \KS(x,y) = \min_{y' \in \calP(n,k)} \KS(x,y'). \]
A mechanism $M$ is 
\emph{KS-optimal}
if for every population $x$,
all (possibly mixed) equilibria 
of the mechanism game corresponding to $M$
assign probability $1$ to KS-optimal samples for $x$.
\end{definition}

The definition of KS-optimality is very strong;
it requires that
\emph{any} sample that may be obtained in an equilibrium of $M$, is not worse (according to the KS statistic)
than \emph{any} other sample.

\bigskip
\noindent\textbf{\textsc{$\W$ Statistic.}} 
The $\W$ statistic compares the population and the sample by measuring the \emph{average} distance between their cumulative distribution functions. Unlike the KS statistic, which focuses on the maximal deviation, the $\W$ statistic aggregates discrepancies across the entire distribution. It is often preferred in settings where noise in the tails is important.


\begin{definition}[$\W$ statistic]
Let $x$ be the population, 
let $\succsim$ be a ranking on $x$, 
and let $y$ be a sample of size $k$ from $x$.
The \emph{$\W$ statistic} of $x$ and $y$ is
\[ \W(x,y) \equiv  \frac{1}{n}\sum_{i=1}^{n} |F_x(x_i) - F_{y}(x_i)|. \]
\end{definition}

Sample and mechanism $\W$-optimality are defined in a similar way to sample and mechanism KS-optimality.

\begin{definition}[$\W$-optimality]
A sample $y$ is \emph{$\W$-optimal} 
for the population $x$ if
\[ \W(x,y) = \min_{y' \in \calP(n,k)} \W(x,y'). \]
A mechanism $M$ is 
\emph{$\W$-optimal}
if for every population $x$,
all (possibly mixed) equilibria 
of the mechanism game corresponding to $M$
assign probability $1$ to $\W$-optimal samples for $x$.\end{definition}

\bigskip
\noindent\textbf{\textsc{Cram\'er-von Mises Statistic.}} 
The Cram\'er-von Mises (CvM) statistic \citep{cramr1928composition, mises2013wahrscheinlichkeit} compares the population and the sample by measuring the \emph{average square} distance between their cumulative distribution functions. It is more sensitive to noise tail than the $\W$ statistic.


\begin{definition}[The CvM statistic]
Let $x$ be the population, 
let $\succsim$ be a ranking on $x$, 
and let $y$ be a sample of size $k$ from $x$.
The \emph{CvM statistic} of $x$ and $y$ is
\[ \CvM(x,y) \equiv  \frac{1}{n}\sum_{i=1}^{n} (F_x(x_i) - F_{y}(x_i))^2. \]
\end{definition}

CvM-optimality is defined in a similar way to KS- and $\W$-optimality.

\begin{definition}[CvM optimality]
A sample $y$ is \emph{CvM-optimal} 
for the population $x$ if
\[ \CvM(x,y) = \min_{y' \in \calP(n,k)} \CvM(x,y'). \]
A mechanism $M$ is 
\emph{$\CvM$-optimal}
if for every population $x$,
all (possibly mixed) equilibria 
of the mechanism game corresponding to $M$
assign probability $1$ to CvM-optimal samples for $x$.\end{definition}

\begin{remark}\label{remark:ordinal}
    Both the population and sample CDFs are ordinal concepts. They do not depend on the specific values that are assigned to the items in the population. If $x=(x_i)_{i=1}^{n}$ and $x'=(x_i')_{i=1}^{n}$ are two different populations with two different rankings $\succsim, ~ \succsim'$, respectively, and if $x_i \succsim x_j$ if and only if $x_i' \succsim' x_j'$, then $F_x(x_i)=F_{x'}(x_i')$ for every $i\in\{1,\ldots,n\}$.
    It therefore follows that our notion of representativeness depends only on the sampled quantiles of the population, and not on the specific values of the sampled items. 

    To illustrate this point, consider two populations that each consist of four natural numbers that are ranked by their order.
The first population is $x = \{1,10,12,100\}$,
and the second population is $\widetilde x = \{1,2,3,4\}$.
The samples $y = \{10,12\}$ and $\widetilde y = \{2,3\}$ consist of the middle two items in the two populations.
Therefore, $\KS(x,y) = \KS(\widetilde x,\widetilde y)$,
$\W(x,y) = \W(\widetilde x,\widetilde y)$,
and
$\CvM(x,y) = \CvM(\widetilde x,\widetilde y)$.

It therefore follows
that whether the sample $y$ represents well the population $x$, does not depend on the absolute values of the smallest and largest (or any other) items in $x$.
\end{remark}

The next mechanism, which we call the \emph{Quantile Mechanism}, is a cut-and-choose mechanism that generates a balanced, or representative sample of the population.

\begin{example}[Quantile mechanism]
\label{example:quantile} 
Fix a population $x = (x_i)_{i=1}^{n}$, a sample size $k$, and a ranking $\succsim$ over $x$. Suppose that $n = (2m+1)k$. The Quantile Mechanism is a cut-and-choose mechanism with one subset with $m+1$ indices, and $k-1$ subsets with $2m+1$ indices each (notice that $m$ indices are not assigned to any subset). The Quantile Mechanism implements the selection of the $(m+1)$-th highest rank item, 
the $m+1 + (2m+1)$-th highest ranked item, and so on up to the $m+1 + (k-1)(2m+1)$-th highest ranked item.
The Quantile Mechanism is symmetric: it outputs the same sample regardless of whether Player I or Player II are chosen as cutter (and Player II, and Player I, respectively, are chosen as chooser). 

An alternative way to implement the Quantile Mechanism is to let Player I (or Player II, this implementation is also symmetric) eliminate $m$ indices, and let Player II (resp., Player I) select one index, then let Player I (resp., II) eliminate $2m$ more indices, and let Player II (resp., I) choose one additional index, and so on with Player I (resp., II) always eliminating $2m$ indices, until $k$ indices are selected.
\end{example}

Our main result establishes the optimality the Quantile mechanism. It shows that the sample obtained by the Quantile mechanism is strictly better than \emph{any} other sample (not equivalent to it) according to all three distance measures introduced above.

\begin{theorem}[Characterization of the optimal mechanism]
\label{theorem:main}
Suppose that $n=(2m+1)k$.
Then, the Quantile mechanism 
is KS-optimal, $\W$-optimal, and CvM-optimal.
Moreover, for every population $x$ and ranking $\succsim$, every sample $y'$ that is not equivalent to the sample $y$ that is produced by the Quantile mechanism $y$
\begin{align*}
&\KS(x,y') > \KS(x,y), \qquad 
\W(x,y') > \W(x,y), \\
\text{and} \qquad 
&\CvM(x,y') > \CvM(x,y).
\end{align*}
\end{theorem}

Theorem~\ref{theorem:main} illustrates the advantage of the Quantile mechanism over \emph{any} other mechanism, whether deterministic or random.
In particular, any deterministic mechanism that does not generate an outcome that is equivalent to the outcome of the Quantile mechanism is \emph{strictly worse} that the Quantile mechanism according to the Kolmogorov-Smirnov statistic, the $\W$ statistic, and the Cram\'er-von Mises statistic.
A random mechanism can never generate an outcome that is better than the Quantile mechanism,
and, if with some positive probability it outputs a sample that is not equivalent to the output of the Quantile mechanism, then it is necessarily worse.

\begin{remark}[The dependence of the statistics on $n$, $m$, and $k$]
For every population $x$, if the ranking $\succsim$ of the items in the population is strict,
then the KS, $\W$, and CvM statistics of the sample $y$ that is produced by the Quantile mechanism are equal to 
\begin{align*}
&\KS(x,y) = \frac{m}{n} = \frac{1}{2k}\left(1-\frac 1n\right), \qquad 
\W(x,y) = \frac{m(m+1)}{n(2m+1)} = \frac{1}{4k}\left(1-\left(\frac kn\right)^2\right), \\
\text{and} \qquad 
&\CvM(x,y) = \frac{2m(m+1)}{n^2} = \frac{1}{2k^2}\left(1-\left(\frac kn\right)^2\right),
\end{align*}
respectively. 
These bounds describe the discrepancy between the sample generated by the Quantile mechanism and the population as a function of the sample size and the population size, 
and describe the rate at which the distance between the sample and population CDFs decreases as a function of the sample and population sizes, $k$ and $n$, respectively.
\end{remark}

The intuition for the proof of Theorem \ref{theorem:main} is that, according to all three measures, the distance between the population and sample CDFs $F_x$ and $F_y$ is maximized right before and right at the chosen quantiles, and decreases on the distance from the closest chosen quantile. Suppose, without loss of generality that $x_1\precsim\cdots\precsim x_n$. Then, right before the first chosen quantile $x_{m+1}$, 
\[
F_x(x_m)-F_y(x_m)=\frac mn -0 = \frac mn.
\]
And, right at the first chosen quantile $x_{m+1}$:
\[
F_x(x_m)-F_y(x_m)=\frac{m+1}{n}-\frac 1k =\frac{m+1}{n}-\frac{2m+1}{n} = -\frac mn,
\]
because the requirement that $n=(2m+1)k$ implies that $\frac 1k = \frac{2m+1}{n}$.
The fact that each additional subset of indices contains $2m+1$ indices ensures that this calculation is repeated right before and right at all other quantiles. For example, right before the second quantile $x_{3m+2}$
\[
F_x(x_{3m+1})-F_y(x_{3m+1})=\frac{3m+1}{n} - \frac 1k = \frac{3m+1}{n} - \frac{2m+1}{n} = \frac mn.
\]
And, right at the second chosen quantile $x_{3m+2}$:
\[
F_x(x_m)-F_y(x_m)=\frac{3m+2}{n}-\frac 2k =\frac{3m+2}{n} - \frac{4m+2}{n} = -\frac mn,
\]
and so on.

The calculation above shows that the Quantile mechanism generates a KS statistic of $\frac mn$. To see why a mechanism that selects different quantiles necessarily generate a strictly larger KS statistic, suppose that the smallest quantile chosen by some alternative mechanism $x_{m'}$ is smaller than the smallest quantile chosen by the Quantile Mechanism, which is given by $x_{m+1}$, or $m'<m+1$. In this case,
\[
F_y(x_{m'})-F_x(x_{m'})=\frac 1k-\frac{m'}{n} = \frac{2m+1}{n}-\frac{m'}{n} > \frac{2m+1}{n} -\frac{m+1}{n}=\frac mn.
\]
And if the statistic $x_{m'}$ is larger than $x_{m+1}$, or $m'>m+1$, then at the point $x_{m'-1}$:
\[
F_x(x_{m'-1})-F_y(x_{m'-1})=\frac{m'-1}{n}- 0 > \frac{m}{n}.
\]
A similar argument shows that if the smallest quantile chosen by an alternative mechanism is equal to $x_{m+1}$ but the next quantile chosen by the mechanism is different from the second quantile chosen by the Quantile Mechanism $x_{3m+2}$, then again the KS statistic generated by the alternative mechanism is strictly larger than $\frac mn$, and do on.

The advantage of the Quantile mechanism is illustrated in Figures~\ref{fig:myfigure1} and~\ref{fig:myfigure2} below.
Figure~\ref{fig:myfigure1} compares the population CDF to the CDF of the sample that is produced by the Quantile mechanism. As can be seen in the figure, 
the sample CDF follows closely the population's CDF,
demonstrating the unique features of the sample produced by the Quantile mechanism -- it is symmetric and it represents well the population.

\begin{figure}[htbp]
  \centering
\includegraphics[width=0.7\textwidth]{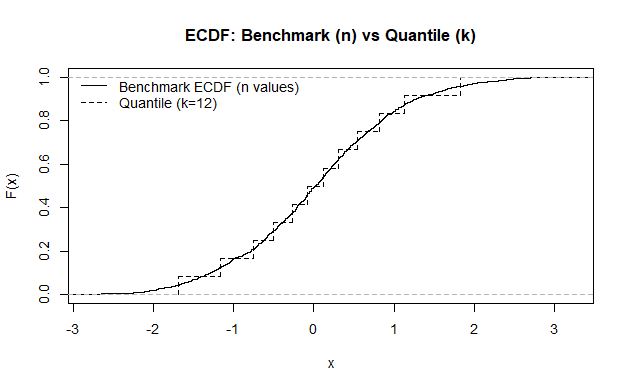}
  \caption{$F_x$ and $F_y$ under the Quantile mechanism for $n=972, k=12, m=40$.}
  \label{fig:myfigure1}
\end{figure}

Figure~\ref{fig:myfigure2} compares the $\KS$ statistic between the population CDF and the sample CDF, for five selection procedures: the Quantile Mechanism, the Random Mechanism (Example \ref{example:uniform}) (with $k$ randomly selected items), the Strike-and-Replace Mechanism (Example \ref{example:struck_jury}), the Median-Sample Mechanism (Example \ref{example:median_sample}), and the Random Mechanism with $k=12$ and $n^*=259$ randomly selected items.\footnote{For the simulations, we draw $972 =(2\cdot 40 +1)\cdot 12$ observations from a standard normal distribution and treat the resulting empirical CDF as the population CDF. The value $n^*=259$ is chosen so that the average KS statistic of a random sample of size $n^*$ from the population CDF matches, up to a small tolerance, the corresponding average distance generated by the Quantile mechanism. To implement the Strike and Replace mechanism, we first draw a random sample of size $k=12$, allow each side to veto up to $3$ observations from opposite ends of the sample, and then refill the struck observations by random draws from the remaining population. To implement the Median-Sample mechanism, we draw $7$ random samples of size $k=12$, rank these samples by their sample medians, and let the two sides veto $3$ samples each from opposite ends, so that the remaining sample is the sample whose median is the median among the candidate sample medians. We repeat this simulation procedure 1,000 times and record the resulting KS statistic for each method.}

As the figure shows, the $\KS$ statistic of the Quantile mechanism is the smallest, and equal to the constant $\frac{m}{n}$.
The $\KS$ statistic of a Cut-and-Choose mechanism in which the player who cuts is required to partition the population into $k$ equally sized subsets is the constant $\frac{2m}{n}$.
The figure also shows the distribution of the $\KS$ statistic of three mechanisms that involve randomization: the median of medians proposed by \cite{flanagan2025not}, the Struck Full Jury, and a random choice of a sample.
These three mechanisms perform significantly worse than the Quantile mechanism.

\begin{figure}[H]
  \centering
\includegraphics[width=0.7\textwidth]{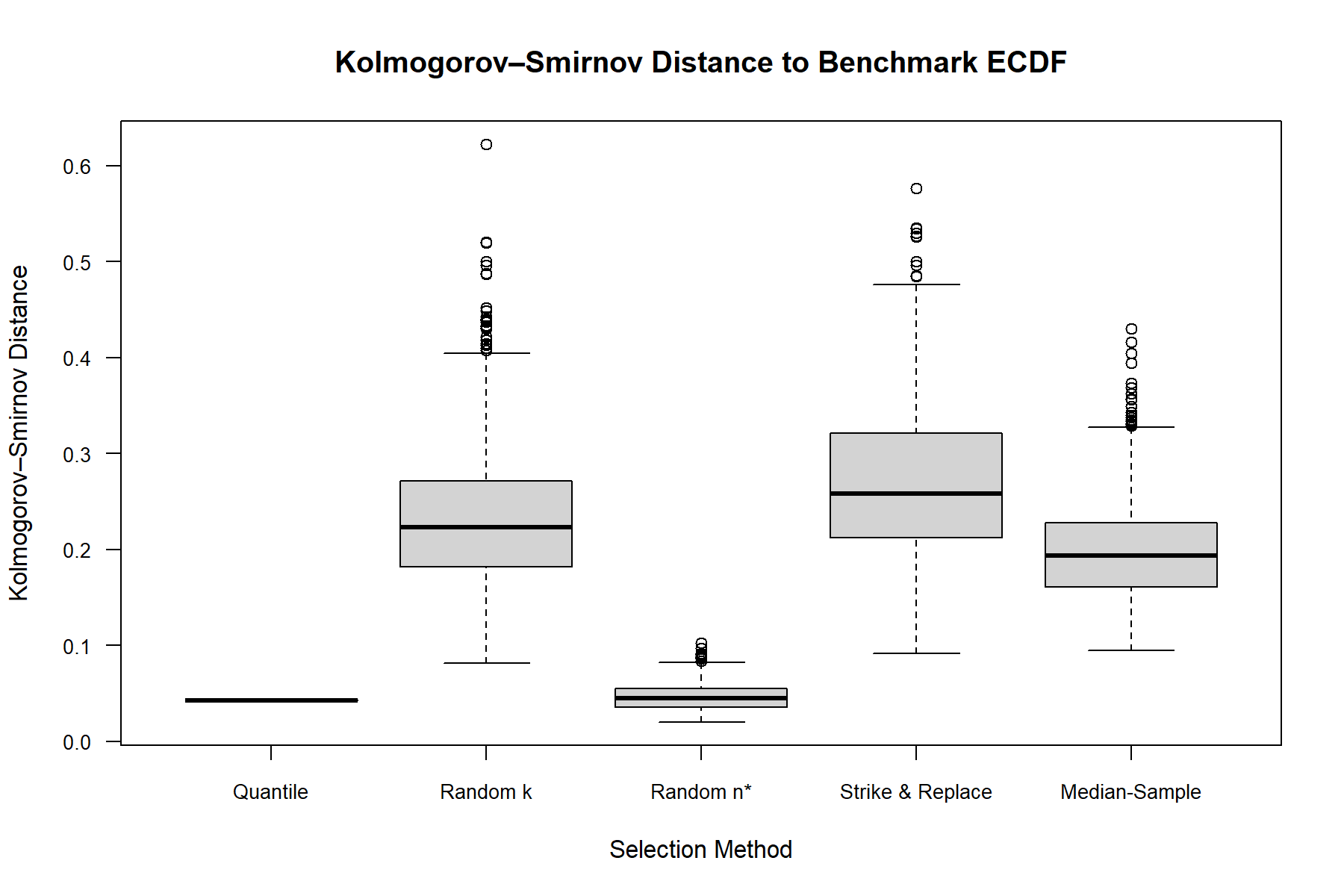}
  \caption{Comparison of the Quantile, Median-Sample, Strike and Replace, and Random mechanisms with $k=12$ and $n^*=259$ chosen items in terms of KS statistic for sampling $100$ times from $n=972, k=12, m=40$.}
  \label{fig:myfigure2}
\end{figure}


\section{Extensions}\label{sec:discussion}

\subsection{Any Selection of Quantiles is Possible}
As mentioned in Remark \ref{remark:ordinal}, the fact that both the population and sample CDFs are ordinal concepts highlights the fact that representativeness depends only on the sampled quantiles of the population, and not on the specific values of the sampled items. 
The next theorem shows that \emph{any selection} of $k$ quantiles can be implemented by some cut-and-choose mechanism.

\begin{theorem}\label{thm:any_selection_is_possible}
    Fix a population size $n$ and a sample size $k\leq n$. Then, any selection of $k$ quantiles $(q_1,\ldots,q_k)\in \{\frac{j_1}{n},\ldots,\frac{j_k}{n}: j_i\in\{1,\ldots,n\} ~ \forall i, j_1<\cdots < j_k\}$ can be implemented by some Cut-and-Choose Mechanism.
\end{theorem}

The proof of Theorem \ref{thm:any_selection_is_possible} follows from the preceding analysis upon observing that the Cut-and-Choose mechanism in which Player II is asked to select $k$ not necessarily disjoint subsets with $q_1,\ldots,q_k$ items each, will lead Player II to select the $k$ subsets with the $q_1$ lowest ranked items, the $q_2$ lowest ranked items and so on, which would induce Player I to select the $q_1$ lowest ranked item, the $q_2$ lowest ranked item, and so on up to the $q_k$ lowest ranked item, as required.

Theorem \ref{thm:any_selection_is_possible} implies that if one believes that the most representative sample consists, say, of the $k$ items which are closest the population sample, then there is a cut-and-choose mechanism that can output these, or any other, quantiles.


\subsection{Non-Antagonistic Players}\label{sec:antagonistic}

So far we assumed that the players are antagonistic:
Player~1 prefers higher-ranked items,
while Player~2 prefers lower-ranked items.
In this subsection we explore the model where each player $i \in \{1,2\}$ has her own ranking $\succsim_i$ on the population,
and the utility function $u_i$ of each player~$i$ is such that
$u_i(y) \geq u_i(y')$ whenever $F_y$ stochastically dominates $F_{y'}$.

The definition of mechanisms and mechanism games remains as before. Since the mechanism game that corresponds to the Quantile mechanism is a game of perfect information (it does not include simultaneous moves), the game admits a subgame-perfect equilibrium in pure strategies. 

The next result shows that
the Quantile mechanism still performs well in the following sense.
For each population $x$, denote by $y(x)$ the outcome of 
a pure subgame-perfect equilibrium of the mechanism game
that is induced by the Quantile mechanism when the two players' preferences are given by $\succsim_1$ and $\succsim_2$, respectively. 
Denote by $y^*_1(x)$ the outcome of the game that is induced by the Quantile mechanism when the ranking of the population is given by Player~1's ranking $\succsim_1$ (that is, when $u_1(y)\geq u_1(y')$ if and only if $F_y$ first-order stochastically dominates $F_{y'}$ and $u_2(y)=-u_1(y)$ for every sample $y$).
And denote by $y^*_2(x)$ the outcome of the game that is induced by the Quantile mechanism
when the ranking of the population is given by Player~2's ranking.

Our result shows that each player $i$ prefers $y(x)$ to $y^*_i(x)$. That is, when players do not have opposing preferences, they both fare better compared to the case in which they each face a player that has opposing preferences to theirs. Intuitively, the player who cuts is better off because they can cut in the same way they would cut if the choosing player had opposing preferences and get an outcome that is weakly better for them from every subset. And the player who chooses is better off because a player who has opposing preferences cuts in the worst possible way for them. The fact that \emph{both players agree} that the outcome is better than what it would have been if they had opposing preferences implies that whatever is lost in terms of pure representativeness, is more than made up by the fact that both players are made better off.

\begin{theorem}\label{thm:nonantagonsistic}
Given a population $x$ and two rankings $\succsim_1$ and $\succsim_2$, we have 
\begin{enumerate}
\item[i.]
$u_1(y(x)) \geq u_1(y^*_1(x))$ under $\succsim_1$.
\item[ii.]
$u_2(y(x)) \geq u_2(y^*_2(x))$ under $\succsim_2$.
\end{enumerate}
\end{theorem}

\subsection{One Player}\label{sec:oneplayer}

So far, with the exception of Section \ref{sec:antagonistic}, we have considered the case in which two players have observed the ranking. What if there is only one player, or if one player does not observe the ranking? 
Such a case may arise, for example, in MDL, if one of the parties holds private information about individual case values.
Consider the following mechanism.

\begin{definition}[Random Cut-and-Choose Mechanism]\label{def:randomC&C}
Suppose $n=km$.
The \emph{Random Cut-and-Choose Mechanism} is the mechanism in which Player~1 who knows the ranking partitions the population into $k$ disjoint subsets, each of size $m$, and one item is randomly chosen from each subset, uniformly and independently among the subsets.
\end{definition}

The outcome of the random Cut-and-Choose mechanism is a random sample $y$. Suppose the items in the population are real numbers,
and the ranking $\succsim$ ranks these numbers by their cardinality. 
Given a partition $P$ selected by Player~1,
the mean $\mu_P(y)$ and the variance $\sigma^2_P(y)$ of the random sample $y$ are random variables,
which depend on the partition.
We show that 
the sample mean $\mu_P(y)$ is independent of the partition,
and the partition that minimizes the variance $\sigma^2_P(y)$ is the equilibrium partition of the Random Cut-and-Choose mechanism
described in Example~\ref{def:randomC&C}.

\begin{theorem}\label{thm:oneplayer}
Suppose that $n = km$. For any partition chosen by the player,
\[
\mu_P(y) = \frac{1}{n}\sum_{i=1}^n x_i .
\]
Moreover, the variance $\E[\sigma^2_P]$ is minimized by the ordered partition, in which the items are assigned to the $k$ subsets according to their order, so that the largest item in each subset is smaller than the smallest item in the next subset.
\end{theorem}

It follows that the player cannot affect the mean of the chosen sample. If the player is risk-averse and prefers the variance of the sample to be as small as possible,  then the random Cut-and-Choose mechanism induces the ordered partition. In this partition, the median items of each subset in the partition coincide with the quantiles selected by the Quantile mechanism, respectively, by their order.

\section{Conclusion}\label{sec:conclusion}

 We consider the problem of how to choose a small representative sample from a large population. As the three legal examples of jury selection, MDL, and redistricting commissions illustrate, current methods for making such a selection are unsatisfactory.
 
 We introduce a selection procedure, the Quantile mechanism, which selects the most representative sample according to three different measures that are prevalent in statistics:  Kolmogorov-Smirnov, $\W$, and Cr\'amer-von Mises. This procedure, which is a symmetrized version of the Cut-and-Choose mechanism, is easy to explain and simple to implement. Indeed, implementing the mechanism requires the  players only to sort the population,  a task whose complexity is $O(n\log(n))$.

The Quantile mechanism strictly outperforms current selection methods, including random selection, and several variants of party selection. 
We envision our findings leading to the adoption of the Quantile mechanism in legal, political, and institutional settings in which adversary parties may influence the choice of samples, sample sizes are tightly constrained, and the overarching social objective is to select a sample that mirrors the population as faithfully as possible.

\bibliographystyle{aea}
\bibliography{bibliography}

\appendix

\section*{Proof of Theorem~\ref{theorem:main}}
\label{sec:proof}

Fix a population $x = (x_1,\dots,x_{n})$ and a ranking $\precsim$ on $x$.
Assume w.l.o.g.~that $x_1 \precsim x_2 \precsim \dots \precsim x_{n}$.
As mentioned in Example~\ref{example:quantile},
when $n=(2m+1)k$,
in all equilibria 
of the quantile mechanism,
the selected sample is equivalent to 
\begin{equation}
    \label{equ:y}
y = (x_{m+1}, x_{m+1+(2m+1)}, \dots, x_{m+1+(2m+1)(k-1)}). 
\end{equation}

We divide the proof into three parts.
In Section~\ref{section:proof:ks} we prove that the quantile mechanism is KS-optimal,
in Section~\ref{section:proof:l1} we prove that the quantile mechanism is $\W$-optimal,
and
in Section~\ref{section:proof:cvm} we prove that the quantile mechanism is $CvM$-optimal.

\subsection*{The quantile mechanism is KS-optimal}
\label{section:proof:ks}

In Step~1 we will show that $\KS(x,y) \leq \frac{m}{n}$,
where $y$ is defined in Eq.~\eqref{equ:y}.
In Step~2 we will show that 
for any sample $y'$ that is not equivalent to $y$,
we necessarily have $\KS(x,y') > \frac{m}{n}$.
Therefore, $\KS(x,y) < \KS(x,y')$.
This in particular holds for any
sample $y'$ which is possible under some equilibrium in the mechanism game that corresponds to some mechanism $M'$.
Since $x$ is arbitrary,
this implies that the quantile mechanism is KS-optimal.

\bigskip
\noindent\textbf{Step 1:
$\KS(x,y) \leq \frac{m}{n}$.}

To prove that
$\KS(x,y) \leq \frac{m}{n}$,
we will show that
\begin{equation}
\label{equ:18}
|F_x(x_i) - F_y(x_i)| \leq \frac{m}{n}, \ \ \ \forall i \in \{1,\dots,n\}.
\end{equation}

Fix $i \in \{1,\dots,n\}$.
We will characterize in turn $F_x(x_i)$ and $F_y(x_i)$.
Recall that we assumed w.l.o.g.~that $x_1 \precsim x_2 \precsim \dots \precsim x_{n}$.

If $x_i \prec x_{i+1}$, then $F_x(x_i) = \frac{i}{n}$.
Otherwise, $x_i \approx x_{i+1}$,
and $F_x(x_i)$ is higher than $\frac{i}{n}$.
In fact, denote by 
$\ell$ the smallest index such that $x_i \prec x_\ell$,
so that
$x_i \approx x_{i+1} \approx \dots \approx x_{\ell-1} \prec x_\ell$.
With this notation, $F_x(x_i) = \frac{\ell-1}{n}$.

We turn to calculate $F_y(x_i)$.
The number of sample elements $y_j$ in $y$ that satisfy $y_j \precsim x_i$ is the maximal $r$ such that 
$m+1+(2m+1)(r-1) < \ell$.
In particular, $\ell \leq m+1+(2m+1)r$ and $F_y(x_i) = \frac{r}{k} = \frac{r(2m+1)}{n}$.
Note that if $r=0$ then $x_i$ is strictly lower than all items in $y$,
while if $r=m$ then $x_i$ is weakly larger than all items in $y$.
Finally,
\begin{align}
-\frac{m}{n}
\nonumber 
&= 
\frac{m+1+(2m+1)(r-1)}{n} - \frac{r(2m+1)}{n}\\
\label{equ:81}
&\leq
\frac{\ell-1}{n} - \frac{r(2m+1)}{n}
=
F_x(x_i) - F_y(x_i)\\
\label{equ:82}
&\leq 
\frac{m+(2m+1)r}{n} - \frac{r(2m+1)}{n}\\
\nonumber
&= \frac{m}{n},
\end{align}
where Eq.~\eqref{equ:81} holds since $m+1+(2m+1)(r-1) < \ell$
and Eq.~\eqref{equ:82} holds since $\ell \leq m+1+(2m+1)r$.
Thus, Eq.~\eqref{equ:18} holds.

\bigskip
\noindent\textbf{Step 2:
$\KS(x,y') > \frac{m}{n}$,
for every 
sample $y'$ that is not equivalent to $y$.}

Fix
a sample $y' = (x_{\ell_1},\dots,x_{\ell_{k}})$
that is not equivalent to $y$.
Assume w.l.o.g.~that $\ell_1 < \dots < \ell_{k}$.
Let $r \in \{1,\dots,k\}$ be the minimal index
such that the $r$'th item in $y'$ differs from the $r$'th item in $y$;
that is,
$x_{\ell_{r}} \not\approx x_{m+1+(2m+1)(r-1)}$.

If $x_{\ell_{r}} \prec x_{m+1+(2m+1)(r-1)}$,
then in particular $\ell_{r} < m+1+(2m+1)(r-1)$ and
$F_x(x_{\ell_{r}}) < \frac{m+1+(2m+1)(r-1)}{n}$.
Moreover, $F_{y'}(x_{\ell_{r}}) \geq \frac{r}{k}$.
Therefore,
\[ F_{y'}(x_{\ell_{r}}) - F_x(x_{\ell_{r}})
> \frac{r}{k} - \frac{m+1+(2m+1)(r-1)}{n} = \frac{m}{n}. \]


If $x_{\ell_{r}} \succ x_{m+(2m+1)(r-1)}$,
then in particular $\ell_{r} > m+(2m+1)(r-1)$.
Hence,
$F_x(x_{m+(2m+1)(r-1)}) > \frac{m+(2m+1)(r-1)}{n}$.
Since $x_{\ell_{r}} \succ x_{m+1+(2m+1)(r-1)}$,
we have $F_y(x_{m+1+(2m+1)(r-1)}) = \frac{r-1}{k}$.
Therefore,
\[ F_x(x_{m+(2m+1)(r-1)}) - F_{y'}(x_{m+(2m+1)(r-1)}) > \frac{m+(2m+1)(r-1)}{n} - \frac{r-1}{k} = \frac{m}{n}. \]
In both cases, $\KS(x,{y'}) > \frac{m}{n}$,
as we wanted to prove.

\subsection*{The quantile mechanism is $\W$-optimal}
\label{section:proof:l1}


We will show that \emph{any} sample $y'$ that is not equivalent to $y$ satisfies
$\W(x,y') > \W(x,y)$.
Since this inequality holds in particular 
for every mechanism $M'$ and every sample $y'$ 
which is possible under some equilibrium of $M'$,
it will follow that the quantile mechanism is $\W$-optimal.

Let then $y' = (x_{\ell_1},x_{\ell_2},\dots,x_{\ell_{k}})$ be any sample that is not equivalent to $y$,
and assume w.l.o.g.~that $\ell_1 < \ell_2 < \dots < \ell_{k}$.

\bigskip
\noindent\textbf{Step 1:
The idea of the proof.}

To prove that $\W(x,y) < \W(x,{y'})$,
we will construct a sequence $y^{(1)},y^{(2)},\dots,y^{(p)}$ of samples such that
$y^{(1)} = y$,
$y^{(p)} = y'$,
and
$\W(x,y^{(\pi)}) < \W(x,y^{(\pi+1)})$
for each $\pi \in \{1,\dots,p-1\}$.

To prove the existence of such a sequence, we will argue that if $y' \not\approx y$,
then there is another set of $k$ indices $(j_1,j_2,\dots,j_{k})$ that satisfies the following properties:
\begin{enumerate}
\item[P1)]
$(j_1,j_2,\dots,j_{k})$ and $(\ell_1,\ell_2,\dots,\ell_{k})$ differ by exactly one index.
\item[P2)]
$(j_1,j_2,\dots,j_{k})$ is closer to $(m+1,m+1+(2m+1),\dots,m+1+(2m+1)(k-1))$ than $(\ell_1,\ell_2,\dots,\ell_{k})$ 
in a sense that will be define shortly.
\item[P3)]
Denoting $y'' = (x_{j_1},x_{j_2},\dots,x_{j_{k}})$,
we have $\W(x,y'') < \W(x,y')$.
\end{enumerate}
A recursive application of this result,
yields a sequence $(y^{(\pi)})_{\pi=1}^p$ as described above.

\bigskip
\noindent\textbf{Step 2:
Defining the sense in which 
$(j_1,j_2,\dots,j_{k})$ is closer to $(m+1,m+1+(2m+1),\dots,m+1+(2m+1)(k-1))$ than $(\ell_1,\ell_2,\dots,\ell_{k})$.}

We will say that 
index $j_r$ is \emph{closer to} $m+1+(2m+1)(r-1)$ than index $l_r$ 
if one of the following conditions holds:
\begin{itemize}
\item $x_{j_r} \approx x_{m+1+(2m+1)(r-1)}$
and $x_{\ell_r} \not\approx x_{m+1+(2m+1)(r-1)}$.
\item $x_{j_r} \not\approx x_{m+1+(2m+1)(r-1)}$,
$x_{\ell_r} \not\approx x_{m+1+(2m+1)(r-1)}$,
and the number of items in $x$ 
that lies strictly between $x_{j_r}$ and $x_{m+1+(2m+1)(r-1)}$ is smaller than the number of items in $x$ 
that lies strictly between
$x_{\ell_r}$ and $x_{m+1+(2m+1)(r-1)}$.
\end{itemize}

By definition, 
it cannot be that \emph{both} $j_r$ is closer to $x_{m+1+(2m+1)(r-1)}$ than $\ell_r$,
and $\ell_r$ is closer to $x_{m+1+(2m+1)(r-1)}$ than $j_r$.

If $j_r$ is not closer to $x_{m+1+(2m+1)(r-1)}$ than $\ell_r$,
and $\ell_r$ is not closer to $x_{m+1+(2m+1)(r-1)}$ than $j_r$,
then either (a) $x_{j_r} \approx x_{\ell_r}$,
or 
(b) $x_{j_r} \not\approx x_{\ell_r}$
and the number of items in $x$ 
that lies strictly between $x_{j_r}$ and $x_{m+1+(2m+1)(r-1)}$ is equal to the number of items in $x$ 
that lies strictly between
$x_{\ell_r}$ and $x_{m+1+(2m+1)(r-1)}$.
Note that if (b) holds while (a) does not,
then necessarily one among $j_r$ and $\ell_r$ is smaller than 
$m+1+(2m+1)(r-1)$ while the other is larger than $m+1+(2m+1)(r-1)$.

We will say that 
the vector $(j_1,j_2,\dots,j_{k})$ is closer to $(m+1,m+1+(2m+1),\dots,m+1+(2m+1)(k-1))$ than the vector $(\ell_1,\ell_2,\dots,\ell_{k})$ if
for the minimal $r$ such that $j_r$ is closer to $m+1+(2m+1)(r-1)$ than index $\ell_r$ 
is smaller than the minimal $r$ such that $\ell_r$ is closer to $m+1+(2m+1)(r-1)$ than index $j_r$ .

\bigskip
\noindent\textbf{Step 3:
Determining the  minimal item in which $y'$ and $y$ differ.}

Let $i$ be the largest index such that $x_{\ell_i} \approx x_{m+1+(2m+1)(i-1)}$.
This is a way to say that 
up to equivalence, 
the samples $y'$ and $y$ agree in the lower $i$ items.
If $x_{\ell_1} \not\approx x_{m+1}$,
we set $i:=0$.

If $i = k$,
then $y' \approx y$.
Since by assumption $y$ and $y'$ are not equivalent,
$i < k$,
and hence $x_{\ell_{i+1}} \not\approx x_{m+1+(2m+1)i}$.
In particular, $\ell_{i+1} \neq m+1+(2m+1)i$.

\bigskip
\noindent\textbf{Step 4: The case $\ell_{i+1} > m+1+(2m+1)i$.}

We will show that in this case it is better to replace $x_{\ell_{i+1}}$ in $y'$ 
by some $x_\ell$ with $\ell < \ell_{i+1}$.

\bigskip
\noindent\textbf{Step 4.1:
Defining the new set of vertices $\vec j$.}

Let $\ell < \ell_{i+1}$ be the largest index such that $x_\ell \prec x_{\ell_{i+1}}$.
In particular,
\[ x_{\ell} \prec x_{\ell+1} \approx x_{\ell+2} \approx \dots \approx x_{\ell_{i+1}}. \]
Since $x_{\ell_{i+1}} \not\approx x_{m+1+(2m+1)i}$,
we have $\ell \geq m+1+(2m+1)i$.

Consider the 
set of indices $\vec j$ that is obtained from $\vec\ell$
by replacing $\ell_{i+1}$ with $\ell$:
\[ \vec j = (\ell_1,\ell_2,\dots,\ell_i,\ell,\ell_{i+2},\dots,\ell_{k}). \]
In particular, (P1) holds.
Since $\vec \ell$ and $\vec j$ coincide in their lower $i+1$ items,
and since $x_{m+1+(2m+1)i} \precsim x_\ell \prec x_{\ell_{i+1}}$,
$\vec j$ is closer to $(m+1,m+1+(2m+1),\dots,m+1+(2m+1)(k-1))$ than $\vec \ell$, and (P2) holds.

Denote by $y''$ the sample induced by the vector of indices $\vec j$:
\[ y'' = (x_{\ell_1},x_{\ell_2},\dots,x_{\ell_i},x_{\ell},x_{\ell_{i+2}},\dots,x_{\ell_{k}}). \]

\noindent\textbf{Step 4.2:
The difference between $\W(x,y')$ and $\W(x,y)$.}

Since $\ell < \ell_{i+1}$,
the statistics $\W(x,{y'})$ and $\W(x,y'')$ differ in the summands that correspond to $x_{\ell}, x_{\ell+1},\dots,x_{\ell_{i+1}-1}$:
since we changed the index $\ell_{i+1}$ by $\ell < \ell_{i+1}$,
we add $\frac{1}{k}$ to the sample's CDF at the points $x_\ell,\dots,x_{\ell_{i+1}-1}$.
The following table describes the quantities $F_x$, $F_{y'}$, and $F_{y''}$ at these points,
where $F_x(x_{\ell_{i+1}}) \geq \frac{\ell_{i+1}}{n}$:
\[
\begin{array}{l|cccc}
& x_\ell & x_{\ell+1} & \dots & x_{\ell_{i+1}-1} \\
\hline
F_x & \frac{\ell}{n} & F_x(x_{\ell_{i+1}}) & \dots & F_x(x_{\ell_{i+1}}) \\
F_{y'} & \frac{i}{k} & \frac{i}{k} & \dots & \frac{i}{k} \\
F_{y''} & \frac{i+1}{k} & \frac{i+1}{k} & \dots & \frac{i+1}{k}
\end{array}
\]
It is thus sufficient to show that 
\begin{equation}
\label{equ:84}
\left|\frac{\ell}{n} - \frac{i}{k} \right| \stackrel?> \left|\frac{\ell}{n} - \frac{i+1}{k}\right|
\end{equation}
and
\begin{equation}
\label{equ:85}
\left|F_x(x_{\ell_{i+1}}) - \frac{i}{k} \right| \stackrel?> \left|F_x(x_{\ell_{i+1}}) - \frac{i+1}{k}\right|.
\end{equation}

\bigskip
\noindent\textbf{Step 4.3: Eq.~\eqref{equ:84} holds when $\frac{\ell}{n} \geq \frac{i+1}{k}$.}

If $\frac{\ell}{n} \geq \frac{i+1}{k}$, then Eq.~\eqref{equ:84} holds.
Indeed, in this case 
we can remove the absolute values from both sides of Eq.~\eqref{equ:84},
and from the right-hand side we subtract a larger amount than from the left-hand side.

\bigskip
\noindent\textbf{Step 4.4: Eq.~\eqref{equ:84} holds when $\frac{\ell}{n} < \frac{i+1}{k}$.}

In this case we should then show that 
\begin{equation}
\label{equ:87}
\frac{\ell}{n} - \frac{i}{k}  \stackrel{?}{>} \frac{i+1}{k} - \frac{\ell}{n}, 
\end{equation} 
which solves to
\begin{equation*}
2\ell \stackrel{?}{>} (2i+1)(2m+1). 
\end{equation*}
However, this inequality holds since $\ell \geq m+1+(2m+1)i$.

\bigskip
\noindent\textbf{Step 4.5: Eq.~\eqref{equ:85} holds when $F_x(x_{\ell_{i+1}}) \geq \frac{i+1}{k}$.}

The argument is similar to that in Step~4.3.

\bigskip
\noindent\textbf{Step 4.6: Eq.~\eqref{equ:85} holds when $F_x(x_{\ell_{i+1}}) < \frac{i+1}{k}$.}

In this case we need to show that 
\begin{equation}
\label{equ:88}
F_x(x_{\ell_{i+1}}) - \frac{i}{k}  \stackrel{?}{>} \frac{i+1}{k} - F_x(x_{\ell_{i+1}}). 
\end{equation}
Since Eq.~\eqref{equ:87} holds,
and since $F_x(x_{\ell_{i+1}}) > \frac{\ell}{n}$,
Eq.~\eqref{equ:88} holds as well.
Indeed, the left-hand side in Eq.~\eqref{equ:88} is larger than the left-hand side in Eq.~\eqref{equ:87},
while the right-hand side in Eq.~\eqref{equ:88} is smaller than the right-hand side in Eq.~\eqref{equ:87}.

\bigskip
\noindent\textbf{Step 5:
The case $\ell_{i+1} < m+1+(2m+1)(i+1)$.}

Note that in this case 
$x_{\ell_{i+1}} \prec x_{m+1+(2m+1)(i+1)}$,
and it may happen that $x_{m+1+(2m+1)(i+1)}$ lies in the sample $y'$.

We will construct $y''$ by replacing $x_{\ell_{i+1}}$ 
with an item $x_\ell$ with $\ell > \ell_{i+1}$.
In fact, $\ell$ will be the smallest index that is larger than $\ell_{i+1}$ and does not appear among the indices that define $y'$.

\bigskip
\noindent\textbf{Step 5.1:
Defining the new set of vertices $\vec j$.}

Let $\ell > \ell_{i+1}$ be the minimal index such that 
(a) $x_\ell \not\approx x_{\ell_{i+1}}$,
and
(b) $\ell \not\in \{\ell_{i+2},\ell_{i+3},\dots,\ell_{k}\}$.
We will show that in this case it is better to replace $x_{\ell_{i+1}}$ in $y'$ 
by $x_\ell$.

We first argue that such an index $\ell$ exists.
Indeed, 
since $m \geq 1$,
and since $\ell_{i+1} < m+1+(2m+1)(i+1)$,
the number of indices in $\{1,\dots,n\}$ larger than $\ell_{i+1}$ is larger than $2(k-i)$.
However, the number of items in $y'$ larger than $x_{\ell_{i+1}}$ is $k-i-1$.
Hence there is at least one index $\ell > \ell_{i+1}$ that satisfies (a) and (b).

Let $\vec j$ be the vector that is derived from $\vec \ell$ by replacing $\ell_{i+1}$ with $\ell$,
so that (P1) holds.
Since $\ell_{i+1} < \ell$, this vector is closer to $(m+1,m+1+(2m+1),\dots,m+1+(2m+1)(k-1))$ than $\vec\ell$ in the lexicographic order,
and (P2) holds. 
We will show that the output vector 
\[ y'' := y' \setminus \{x_{\ell_{i+1}}\} \cup \{x_\ell\} \]
satisfies 
$\W(x,{y'}) > \W(x,y'')$,
so that (P3) holds as well.

\bigskip
\noindent\textbf{Step 5.2:
The difference between $\W(x,y')$ and $\W(x,y)$.}

Since $\ell_{i+1} < \ell$,
the statistics $\W(x,{y'})$ and $\W(x,y'')$ differ in the summands that correspond to 
all $r$ with $x_{r} \approx x_{\ell_{i+1}}$,
and in the summands that correspond to 
$\ell_{i+1}+1$, $\ell_{i+1}+2$,$\dots$,$x_{\ell-1}$.
Since to create $y''$ from $y'$ we changed $x_{\ell_{i+1}}$ to $x_\ell$, and since $\ell > \ell_{i+1}$,
we have $F_{y'}(x_{j}) = F_y(x_j) + \frac{1}{k}$ for any $j$ such that $x_j \approx x_{\ell_{i+1}}$ and for $\ell_{i+1}+1$, $\ell_{i+1}+2$,$\dots$,$x_{\ell-1}$.
The following table describes the quantities $F_x$, $F_{y'}$, and $F_{y''}$ in these summands:
\[
\begin{array}{l|ccccc}
& x_r \approx x_{\ell_{i+1}} & x_{\ell_{i+1}+1} & x_{\ell_{i+1}+2} & \dots & x_{\ell-1} \\
\hline
F_x(\cdot)& F_x(x_{\ell_{i+1}}) & F_x(x_{\ell_{i+1}+1}) & F_x(x_{\ell_{i+1}+2}) & \dots & F_x(x_{\ell}) \\
F_{y'}(\cdot) & F_{y''}(x_{\ell_{i+1}})+\frac{1}{k} & F_{y''}(x_{\ell_{i+1}+1})+\frac{1}{k} & F_{y''}(x_{\ell_{i+1}+2})+\frac{1}{k} & \dots &
F_{y''}(x_{\ell-1})+\frac{1}{k}  \\
F_{y''}(\cdot) & F_{y''}(x_{\ell_{i+1}}) & F_{y''}(x_{\ell_{i+1}+1}) & F_{y''}(x_{\ell_{i+1}+2}) & \dots &
F_{y''}(x_{\ell-1})
\end{array}
\]

We will show that in \emph{all} these summands, 
the contribution to $\W(x,{y''})$ is smaller than the contribution to $\W(x,y')$.

\bigskip
\noindent\textbf{Step 5.3:
Comparing summands.}

Take any $q \geq 0$ such that $\ell_{i+1}+q < \ell$.
We will show that 
\begin{equation}
\label{equ:14}
\left|F_x(x_{\ell_{i+1}+q}) - F_{y''}(x_{\ell_{i+1}+q}) - \frac{1}{k} \right| \stackrel{?}{>} \left|F_x(x_{\ell_{i+1}+q}) - F_{y''}(x_{\ell_{i+1}+q})\right|.
\end{equation}
If $F_x(x_{\ell_{i+1}+q}) \leq F''(x_{\ell_{i+1}+q})$,
then Eq.~\eqref{equ:14} as in Step~4.3.

Assume then that $F_x(x_{\ell_{i+1}+q}) > F''(x_{\ell_{i+1}+q})$.
To prove Eq.~\eqref{equ:14} it is sufficient to show that
\begin{equation}
\label{equ:71}
F_{y''}(x_{\ell_{i+1}+q}) + \frac{1}{k}  - F_x(x_{\ell_{i+1}+q}) \stackrel{?}{>} F_x(x_{\ell_{i+1}+q}) - F_{y''}(x_{\ell_{i+1}+q}).
\end{equation}
We first relate 
$F_x(x_{\ell_{i+1}+q})$ to $F_x(x_{\ell_{i+1}})$,
and 
$F_{y''}(x_{\ell_{i+1}+q})$ to $F_{y''}(x_{\ell_{i+1}})$.
Let $D \geq 0$ be the number of indices $j$ such that $x_{\ell_{i+1}} \prec x_j \precsim x_{\ell_{i+1}+q}$.
Then 
\begin{equation}
\label{equ:72}
F_x(x_{\ell_{i+1}+q}) = F_x(x_{\ell_{i+1}}) + \frac{D}{n}.
\end{equation}
Since by the definition of $\ell$, for every such $j$, 
$x_j$ is in $y''$, we have
\begin{equation}
\label{equ:73}
F_{y''}(x_{\ell_{i+1}+q}) = F_{y''}(x_{\ell_{i+1}}) + \frac{D}{k}.
\end{equation}
Substituting Eqs.~\eqref{equ:72} and~\eqref{equ:73} in Eq.~\eqref{equ:71} and simplifying the resulting equation, we obtain that we need to show that
\begin{equation}
\label{equ:74}
2F_{y''}(x_{\ell_{i+1}}) + \frac{2D+1}{k}  \stackrel{?}{>} 2F_x(x_{\ell_{i+1}}) + \frac{2D}{n}.
\end{equation}
However, $F_{y''}(x_{\ell_{i+1}}) \geq \frac{i}{k}$,
and $F_x(x_{\ell_{i+1}}) \leq m+(2m+1)i$.
Hence the left-hand side in Eq.~\eqref{equ:74} is
$\frac{(2i+2D+1)(2m+1)}{n}$,
while the right-hand side is at most 
$\frac{2m+2i(2m+1) + 2D}{n}$,
and the former is always larger than the latter.

\subsection*{The quantile mechanism is $CvM$-optimal}
\label{section:proof:cvm}

The proof is analogous to that presented in Section~\ref{section:proof:l1}.
We here detail the differences.

\bigskip
\noindent\textbf{Step 1: The case $\ell_{i+1} > m + (2m+1)(i+1)$.}

When $\ell_{i+1} > m + (2m+1)(i+1)$,
the analog of Eq.~\eqref{equ:84} is
\begin{equation*}
\left(\frac{\ell}{n} - \frac{i}{k} \right)^2 \stackrel?> \left(\frac{\ell}{n} - \frac{i+1}{k}\right)^2.
\end{equation*}
This equation reduces to
\[ 2\frac{\ell}{nk} \stackrel?> \left(\frac{i}{k}+\frac{1}{k}\right)^2 - \left(\frac{i}{k}\right)^2, \]
which solves to
\[ 2\ell \stackrel?> (2i+1)(2m+1). \]
However, this equation holds since $\ell \geq m+1+(2m+1)i$.

\bigskip
\noindent\textbf{Step 2: The case $\ell_{i+1} < m + (2m+1)(i+1)$.}

The analog of Eq.~\eqref{equ:14} is
\begin{equation*}
\left(F_x(x_{\ell_{i+1}+q}) - F_{y''}(x_{\ell_{i+1}+q}) - \frac{1}{k} \right)^2 \stackrel{?}{>} \left(F_x(x_{\ell_{i+1}+q}) - F_{y''}(x_{\ell_{i+1}+q})\right)^2,
\end{equation*}
which simplifies to
\begin{equation}
\label{equ:76}
\frac{1}{k^2} \stackrel{?}{>} \frac{2}{k}
\left( F_x(x_{\ell_{i+1}+q}) - F_{y''}(x_{\ell_{i+1}+q})\right).
\end{equation}
Multiplying both sides of Eq.~\eqref{equ:76} by $k^2$
and 
taking into account Eqs.~\eqref{equ:72} and~\eqref{equ:73},
Eq.~\eqref{equ:76} translates to
\begin{equation*}
1 \stackrel?> 2k\left( F_x(x_{\ell_{i+1}}) - F_{y''}(x_{\ell_{i+1}}) + \frac{D}{n} - \frac{D}{k}\right), 
\end{equation*}
which is equivalent to
\begin{equation*}
2kF_{y''}(x_{\ell_{i+1}}) + 2D +1 \stackrel?> 2kF_x(x_{\ell_{i+1}}) + \frac{2kD}{n}.
\end{equation*}
However, this inequality is equivalent to Eq.~\eqref{equ:74}.

\section*{Proof of Theorem \ref{thm:nonantagonsistic}}

We start by proving Claim (i).
For every partition $P \in \calP(n,k)$,
let $y(P)$ be the sample that contains, 
for each element in $P$, an item that is maximal according to $\precsim_2$.
The sample $y(P)$ is the outcome when Player~1 partitions $x$ according to $P$, Player~2's ranking is $\precsim_2$,
and Player~2 best responds to $P$.

Let $P^* \in \calP(n,k)$ be Player~1's optimal partition under $\precsim_1$.
Note that $y(P^*)$ stochastically dominates $y^*_1$.
Indeed, both samples are derived from the partition $P^*$;
however, $y^*_1$ is derived assuming Player~2's ranking is $\succsim_1$ --
the worst from Player~1's point of view.
Hence, $u_1(y(P^*)) \geq u_1(y^*_1)$ under $\precsim_1$.
Therefore, in equilibrium, Player~1's utility is at least $u_1(y(P^*))$, and (i) follows.

We turn to prove Claim (ii).
We will show that $F_{y^*_2}$ stochastically dominates $F_{y(x)}$ under $\precsim_2$, from which (ii) will follow.

Fix then $i \in \{1,\dots,n\}$,
and suppose that 
$\#\{j \in \{1,\dots,n\} \colon x_j \precsim_2 x_i\} = r$.
In the sample $y^*_2(x)$, there are $\lceil\frac{ r-m }{2m+1}\rceil$ items that are lower than or equivalent to $x_i$ under $\precsim_2$.

Whatever be the partition of Player~1 under the equilibrium, 
the $r$ items in $\{x_j \colon x_j \precsim_2 x_i\}$
lie in at least $\lceil\frac{r}{2m+1}\rceil$ elements of the partition.
Since in equilibrium Player~2 selects the minimal item according to $\precsim_2$ from each element of the partition, there are at least $\lceil\frac{r}{2m+1}\rceil$ items in $y(x)$ that are lower than or equivalent to $x_i$ under $\precsim_2$.
Since $\lceil\frac{r}{2m+1}\rceil \geq \frac{\lceil r-m \rceil}{2m+1}$, we have $F_{y(x)}(x_i) \geq F_{y^*_2(x)}$ according to $\precsim_2$.
Since $i$ is arbitrary, $F_{y^*_2}$ stochastically dominates $F_{y(x)}$, as claimed.

\section*{Proof of Theorem \ref{thm:oneplayer}}

Fix for a moment a partition $P = (Q_1,\dots,Q_k) \in \calP(n,k)$ selected by Player~1.
Since all elements of $P$ include $m$ items,
the probability that item $x_i$ is selected to the sample is $\frac{1}{m}$.
By the linearity of the expectation operator,
\[ \mu_P(y) 
= \frac{1}{k}\sum_{\ell=1}^k\left(\frac{1}{m}\sum_{x_i\in Q_\ell} x_i\right)
= \frac{1}{n}\sum_{i=1}^n x_i. \]
Denote by $(j_1,\dots,j_k)$ the random indices of the items in the sample; that is, $j_\ell$ is selected uniformly from $Q_\ell$, for every $1 \leq \ell \leq k$.
The sample's variance is
\begin{align*}
\sigma^2_P(y) &= \sum_{P \in \calP(n,k)} \left( \frac{1}{k} \sum_{\ell=1}^k x_{j_\ell}^2 -
\left( \frac{1}{k} \sum_{\ell=1}^k x_{j_\ell}\right)^2\right).\end{align*}
Because there are $m^k$ different samples,
and they are all equally likely,
the expected variance is
\begin{align*}
\E[\sigma^2_P(y)]
&= \frac{1}{n} \sum_{i=1}^n x_{i}^2 -
\frac{1}{m^k} \sum_{P \in \calP(n,k)} 
\left(\frac{1}{k} \sum_{\ell=1}^k x_{j_\ell}\right)^2.
\end{align*} 
Since the first summand is independent of $P$,
to minimize $\E[\sigma^2_P(y)]$
it is sufficient to maximize the second summand.
Since the function $x \mapsto x^2$ is convex,
by Jensen's inequality,
the second summand is maximized when $P$ is the partition of the population into $k$ blocks of size $m$,
where each block contains consecutive items according to $\precsim$.

\end{document}